\documentclass[a4paper,fleqn]{cas-dc}

\usepackage[authoryear,longnamesfirst]{natbib}
\usepackage[authoryear]{natbib}
\usepackage{pifont}
\newcommand{\xmark}{\ding{55}}%

\def\tsc#1{\csdef{#1}{\textsc{\lowercase{#1}}\xspace}}
\tsc{WGM}
\tsc{QE}
\tsc{EP}
\tsc{PMS}
\tsc{BEC}
\tsc{DE}

\begin{document}
\let\WriteBookmarks\relax
\def\floatpagepagefraction{1}
\def\textpagefraction{.001}
\shorttitle{Time-domain deep attractor network}
\shortauthors{Hangting Chen et~al.}

\title [mode = title]{Exploring the time-domain deep attractor network with two-stream architectures in a reverberant environment}



\author[1,2]{Hangting Chen}[style=chinese,
							orcid=0000-0002-4085-4364]
\ead{chenhangting@hccl.ioa.ac.cn}


\author[1,2]{Pengyuan Zhang}[style=chinese]
\cormark[1]
\ead{zhangpengyuan@hccl.ioa.ac.cn}


\address[1]{Key Laboratory of Speech Acoustics \& Content Understanding, Institute of Acoustics, CAS, China}

\address[2]{University of Chinese Academy of Sciences, Beijing, China}

\cortext[cor1]{Corresponding author}

\begin{abstract}
Deep attractor networks (DANs) perform speech separation with discriminative embeddings and speaker attractors. Compared with methods based on the permutation invariant training (PIT), DANs define a deep embedding space and deliver a more elaborate representation on each time-frequency (T-F) bin. However, it has been observed that the DANs achieve limited improvement on the signal quality if directly deployed in a reverberant environment. Following the success of time-domain separation networks on the clean mixture speech, we propose a time-domain DAN (TD-DAN) with two-streams of convolutional networks, which efficiently perform both dereverberation and separation tasks under the condition of a variable number of speakers. The speaker encoding stream (SES) of the TD-DAN is trained to model the speaker information in the embedding space. The speech decoding stream (SDS) accepts speaker attractors from the SES and learns to estimate early reflections from the spectro-temporal representations. Meanwhile, additional clustering losses are used to bridge the gap between the oracle and the estimated attractors. Experiments were conducted on the Spatialized Multi-Speaker Wall Street Journal (SMS-WSJ) dataset. The early reflection was compared with the anechoic and reverberant signals and then was chosen as the learning targets. The experimental results demonstrated that the TD-DAN achieved scale-invariant source-to-distortion ratio (SI-SDR) gains of $9.79/7.47$ dB on the reverberant $2/3$-speaker evaluation set, exceeding the baseline DAN and convolutional time-domain audio separation network (Conv-TasNet) by $1.92/0.68$ dB and $0.91/0.47$ dB, respectively.
\end{abstract}


\maketitle

\section{Introduction}

Speech signals captured by distant microphones often present with reverberation, noise and multiple speakers, rendering low speech intelligibility for human listeners. In such situations, obtaining the single-speaker close-talk signal requires the ability to perform dereverberation and source separation, with noise being viewed as a particular source. 



Despite the great success of speech separation on clean close-talk utterances, blind source separation remains challenging in a reverberant environment. Some researchers have designed more sophisticated network architectures by directly mapping the reverberant signals to anechoic signals (\cite{Nachmani2020VoiceSW}\cite{Shi2020LaFurcaIR}). Some studies have performed dereverberation and separation with tandem systems, each part of which is designed for a single task. The framework in (\cite{Nakatani2020DNNsupportedMC}) integrates deep learning-based speech separation, statistical model-based dereverberation and beamforming. Another study (\cite{Maciejewski2019WHAMRNA}) cascades networks to learn different targets and outperforms the spectral mapping from the reverberant mixture to the anechoic signal. \cite{Fan2020SimultaneousDA} proposed deep embedding methods to capture the difference between the anechoic and the residual reverberant signals, which inspired us to train discriminative embeddings for both speaker separation and dereverberation under a unified architecture.


More recently, the time-domain audio separation network (TasNet) has provided a novel separation scheme that works on time-domain representations with a time-domain convolutional encoder and decoder (\cite{Luo2018TasNetSI}). The subsequent Conv-TasNet (\cite{Luo2019ConvTasNetSI}) and other works (\cite{Shi2019FurcaxEM,Bahmaninezhad2019ACS}) have demonstrated significant separation performance that even exceeds that of the ideal time-frequency (T-F) masks. The classic Conv-TasNet uses permutation invariant training (PIT) to generate enhanced signals from different speakers. On the other hand, the deep attractor network (DAN) (\cite{Luo2018SpeakerIndependentSS}) presents another paradigm, which calculates the masks with deep embedding features. Compared with PIT-based methods, the output of DAN forms a deep embedding space and delivers a more elaborate representation on each T-F bin. However, the original DAN is trained in the T-F domain under clean mixture signals, which limits its performance and application under reverberant environments.

In this study, we propose a novel time-domain DAN (TD-DAN) to simultaneously perform dereverberation and separation tasks. The designed architecture consists of $2$ parallel streams, a speaker encoding stream (SES) for speaker embedding modelling and a speech decoding stream (SDS) for speech separation and dereverberation. The SES is trained with a reconstruction loss and clustering losses, resulting in speaker embeddings that are discriminative and suitable for clustering. Moreover, the SDS serves as an inference module that first models the deep embeddings on the spectro-temporal representations and then interacts with the SES to generate enhanced signals. The proposed scheme makes the following contributions: 
\begin{itemize}
	\item Different learning targets are compared using the clean signal as the reference. We have demonstrated that the early reflection is a favorable choice for models to learn the mapping from reverberant mixtures to dereverberated single-speaker signals.
	\item The DAN is extended to the time domain with a two-stream architecture, which generates the embeddings defined on the spectro-temporal representations and performs dereverberation and separation simultaneously. On the 2/3-speaker evaluation (Eval.) set, the TD-DAN achieved scale-invariant source-to-distortion ratios (SI-SDRs) exceeding the DAN and Conv-TasNet by $1.92/0.68$ dB and $0.91/0.47$ dB, respectively.
	\item Clustering losses are employed to bridge the gap between the oracle attractor and K-means clustering under the reverberant environment. 
\end{itemize}

The rest of the paper is organized as follows. In Section 2, we briefly introduce the techniques related to the proposed method. In Section 3, we describe the proposed TD-DAN and the auxiliary clustering loss. Section 4 presents and discusses the experimental results of the proposed methods. Section 5 concludes this work.

\section{Related work}

Previous work on far-field speech separation focused on the following $3$ issues: dereverberation, speech separation and unified frameworks.

\textbf{Dereverberation:} Many algorithms have been proposed, such as beamforming (\cite{Schwartz2016JointML,Kodrasi2017EVDbasedMD,Nakatani2019MaximumLC}) and blind inverse filtering (\cite{Schmid2012AnEA, Yoshioka2012GeneralizationOM}), to address the dereverberation problem. The weighted prediction error (WPE) was developed under the paradigm of blind inverse filtering, which rose to prominence in the REverberant Voice Enhancement and Recognition Benchmark (REVERB) challenge (\cite{Kinoshita2016ASO}). It aims to minimize the prediction error by optimizing the delayed linear filters to eliminate the detrimental late reverberation (\cite{Nakatani2010SpeechDB}).  Deep neural networks (DNNs) have been used to learn the spectral mapping from reverberant signals to anechoic signals (\cite{Geetha2017LearningSM}). In practice, mask estimation is preferred for its superior performance compared with the spectral mapping (\cite{Wang2014OnTT}). Moreover, complex ideal ratio masks (cIRMs) are proposed to overcome the drawback that real-valued masks cannot reconstruct the phase information of the target signal (\cite{Williamson2017TimeFrequencyMI}). Some researchers attempt to combine DNNs with WPE by deep learning-based energy variance estimation, leading to a non-iterative WPE algorithm (\cite{Heymann2019JointOO}).

\textbf{Paradigms of speech separation:} Most architectures adopt $2$ paradigms, PIT and embedding clustering-based methods. PIT (\cite{Kolbaek2017MultitalkerSS}) directly optimizes the reconstruction loss with possible permutations. PIT can deal with the condition of variable speakers by iterative separation (\cite{Takahashi2019RecursiveSS}), model selection (\cite{Nachmani2020VoiceSW}) or assuming a maximum number of sources (\cite{Luo2020SeparatingVN}). Speaker clustering methods such as deep clustering (DC) (\cite{Hershey2016DeepCD,Wang2018AlternativeOF}) are trained to generate discriminative deep embeddings on each T-F bin, and use clustering algorithms to obtain speaker assignment during the test phase. The DAN is developed following DC, but it directly optimizes the reconstruction of the spectrogram (\cite{Luo2018SpeakerIndependentSS}). DC and DANs can deal with a variable number of speakers by setting the cluster number.

\textbf{Learning objects of speech separation:} Most previous approaches have been formulated by predicting T-F masks of the mixture signal. Commonly used masks are ideal binary masks (IBMs), ideal ratio masks (IRMs) and Wiener filter-like masks (WFMs) (\cite{Wang2014OnTT}). Some approaches directly predict the spectrogram of each source (\cite{Du2016ARA}). Both mask estimation and spectrum prediction use the inverse short-time Fourier transform (iSTFT) of the estimated magnitude spectrogram of each source together with the original or the modified phase. Recently, TasNet have introduced a novel method of separating signals from the raw waveform. It utilizes $1$-D convolutional filters to encode the raw waveform and decode the generated spectro-temporal representations. A speech separation module accepts the representation and predicts source masks. Unlike the fixed weights of the short-time Fourier transform (STFT), TasNet learns the transformation weight by optimizing SI-SDRs between the estimated and target source signals.

\textbf{Unified frameworks:} Speech separation in a reverberant environment is a difficult task by simultaneously addressing the dereverberation and separation problems. Some systems adopt algorithms in tandem, for example, the framework in (\cite{Nakatani2020DNNsupportedMC}) combines weighted power minimization distortion-less response (WPD) (\cite{Nakatani2019AUC}), noisy complex Gaussian mixture Model (noisyCGMM) (\cite{Ito2018NoisyCC}), and convolutional neural network (CNN)-based PIT. A purely deep learning-based network is introduced for denoising and dereverberation by learning the noise-free deep embeddings firstly and then performing mask-based dereverberation (\cite{Fan2020SimultaneousDA}). The Conv-TasNet achieved a low SI-SDR to perform both the dereverberation and separation tasks, compared with its performance on the clean WSJ0-2MIX dataset (\cite{Maciejewski2019WHAMRNA}). Some researchers have designed sophisticated architectures and modules to improve the performance (\cite{Nachmani2020VoiceSW}\cite{Shi2020LaFurcaIR}). \cite{Zeghidour2020WavesplitES} proposes a clustering method to capture the long-term representation, but the number of speakers is fixed and the separation is conducted by feature-wise linear modulation (\cite{Perez2018FiLMVR}) instead of the similarity between deep embeddings.

\section{Methods}

In this section, we first formulate the problem and introduce the baseline DAN and Conv-TasNet. Following the design of the speaker attractor and time convolutional network (TCN), 2 types of two-stream TD-DANs are proposed, one with hybrid encoders and another with fully time-domain waveform encoders. Additionally, clustering losses are proposed to improve the performance of the attractors obtained by the K-means clustering algorithm.

\subsection{Problem formulation}

Assume that speech signals from $K$ speakers are captured by a distant microphone in a noisy reverberant environment. The captured signal is
\begin{flalign}
\label{eq:rsig}
y=\sum_{k=1}^{K}y^{(k)}+n=\sum_{k=1}^{K}d^{(k)}+\sum_{k=1}^{K}r^{(k)}+n, &&
\end{flalign}
where $n$ is the noise and $y^{(k)}$ is the reverberant source signal, which is decomposed as $d^{(k)}$ representing the direct sound and early reflection and $r^{(k)}$ representing the late reverberation. For simplicity, $d^{(k)}$ is referred to as the early reflection in the rest of the paper. The STFT transforms the signal to T-F representations, reformulating Eq.(\ref{eq:rsig}) as
\begin{flalign}
y_{t,f}=\sum_{k=1}^{K}y_{k,t,f}+n_{t,f}=\sum_{k=1}^{K}d_{k,t,f}+\sum_{k=1}^{K}r_{k,t,f}+n_{t,f}, &&
\end{flalign}
with $T$ frames, maximum frequency index $F$, frame index $t=1,...,T$ and frequency index $f=0,...,F$. The early reflection $d_{k,t,f}$ and the late part $r_{k,t,f}$ are generated by convolution,
\begin{flalign}
d_{k,t,f}=\sum_{\tau=0}^{D-1}a_{k,\tau,f}s_{k,t-\tau,f}, && \\
r_{k,t,f}=\sum_{\tau=D}^{L_a-1}a_{k,\tau,f}s_{k,t-\tau,f}, &&
\end{flalign}
where $a_{k,f}=[a_{k,0,f},a_{k,1,f},...,a_{k,L_a-1,f}]$ is the transfer function with late reverberation starting from frame $D$ and ending at frame $L_a$ for frequency $f$, and $s_{k,t,f}$ is the source signal for speaker $k$ on bin $t,f$. As indicated in (\cite{Bradley2003OnTI}), the early reflections increase the speech intelligibility scores for both impaired and non-impaired listeners. Moreover, it is indicated in Section \ref{sec:learningtargets} that the early reflection is a favorable learning target for networks to conduct the dereverberation and separation tasks. Thus, in this study, dereverberation is to eliminate the late part $r_{k,t,f}$.

The ideal masks are defined in the T-F domain. The IRM for speech separation only is expressed as
\begin{flalign}
\label{eq:irm1}
m^{\text{IRM(sepr)}}_{k,t,f}=\frac{|y_{k,t,f}|}{\sum_{k}|y_{k,t,f}|+|n_{t,f}|}, &&
\end{flalign}
where $|\cdot|$ is a modulus operation. In the reverberant environment, the IRM for the dereverberated source $k$ is defined as
\begin{flalign}
\label{eq:irm2}
m^{\text{IRM(sepr+derevb)}}_{k,t,f}=\frac{|d_{k,t,f}|}{|y_{t,f}-d_{k,t,f}|+|d_{k,t,f}|}, &&
\end{flalign}
where the interference signal is obtained by removing the early part of source $d_{k,t,f}$, i.e., it includes both the late reverberation of the target source and other interference signals. Similarly, WFM is formulated as
\begin{flalign}
\label{eq:wfm1}
m^{\text{WFM(sepr)}}_{k,t,f}=\sqrt{\frac{|y_{k,t,f}|^2}{\sum_{k}|y_{k,t,f}|^2+|n_{t,f}|^2}}, &&
\end{flalign}
\begin{flalign}
\label{eq:wfm2}
m^{\text{WFM(sepr+derevb)}}_{k,t,f}=\sqrt{\frac{|d_{k,t,f}|^2}{|y_{t,f}-d_{k,t,f}|^2+|d_{k,t,f}|^2}}. &&
\end{flalign}

\subsection{Baseline DAN and Conv-TasNet}

\begin{figure}
	\centering
	\includegraphics[width=\linewidth]{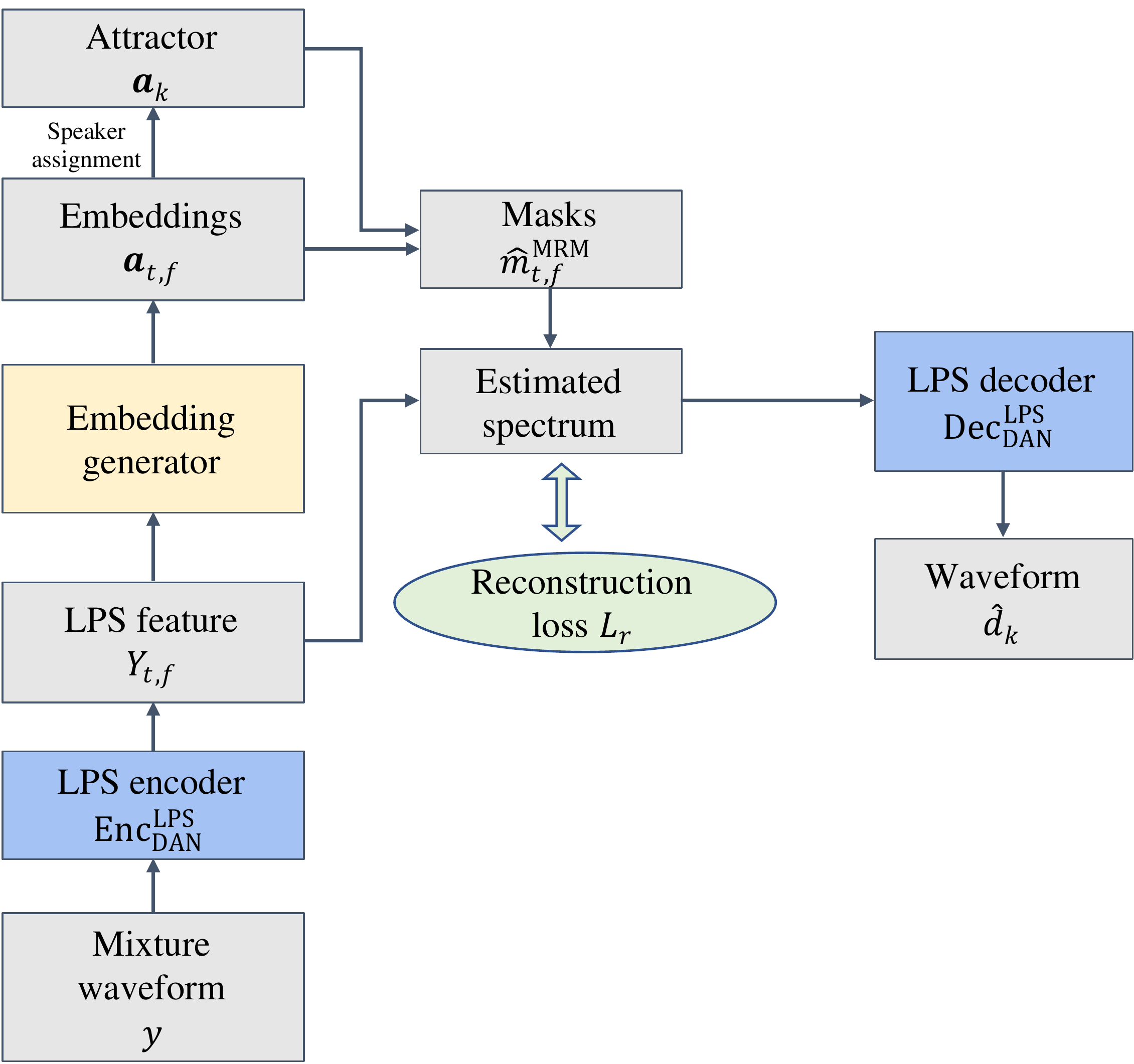}
	\caption{The architecture of the DAN, where the attractor is obtained by oracle assignment and K-means clustering in the training and testing phases, respectively.}
	\label{fig:dan}
\end{figure}

Our TD-DAN is inspired by the design of the deep embedding and TCN, which is originally proposed in DAN (\cite{Luo2018SpeakerIndependentSS}) and Conv-TasNet (\cite{Luo2019ConvTasNetSI}), respectively. We briefly introduce these 2 networks in this section.

\subsubsection{Deep attractor network}

The attractor is a speaker embedding indicating speaker information. As shown in Fig.\ref{fig:dan}, the DAN accepts the log power spectrum (LPS) and generates $D$-dimensional speaker embeddings $\mathbf{a}_{t,f}$,
\begin{flalign}
\label{eq:DAN}
\{\mathbf{a}_{t,f}\}_{t,f}=\text{DAN}(\text{Enc}_\text{DAN}^\text{LPS}(y)), &&
\end{flalign}
where $\{\cdot\}_{\{\cdot\}}$ denotes the matrix form with subscripts representing the axes and $\text{Enc}_{\{\cdot\}}^\text{LPS}$ is the LPS feature extractor. During training, the attractor vector $\mathbf{a}_k$ for speaker $k$ is obtained by averaging over the T-F bins,
\begin{flalign}
\label{eq:attractor}
\mathbf{a}_k=\frac{\sum_{t,f}m^\text{IBM}_{k,t,f}v_{t,f}\mathbf{a}_{t,f}}{\sum_{t,f}m^\text{IBM}_{k,t.f}v_{t,f}}, &&
\end{flalign}
where $v_{t,f}\in \{0,1\}$ denotes the absence/presence of speech calculated by a threshold of power and $m^\text{IBM}_{k,t,f}$ is the binary speaker assignment. Here, we use early reflections to calculate $m^\text{IBM}_{k,t,f}$:
\begin{flalign}
\label{eq:attract_ibm}
m^\text{IBM}_{k,t,f}=
\begin{cases}
0, \text{if $|d_{k,t,f}|\leqslant\sum_{q\neq k}|d_{q,t,f}|$} \\
1, \text{if $|d_{k,t,f}|>\sum_{q\neq k}|d_{q,t,f}|$}
\end{cases}, &&
\end{flalign}
where $\mathbf{a}_{t,f}$ is expected to indicate the source information and can be used to perform both separation and dereverberation. During the testing phase, the attractors are obtained by K-means clustering with prior knowledge of the number of speakers,
\begin{flalign}
\{\mathbf{a}_k\}_k=\text{KMeans}(\{\mathbf{a}_{t,f}|\text{if }v_{t,f}=1\}). &&
\end{flalign} 
The masks are estimated with Sigmoid activation,
\begin{flalign}
\hat{m}^\text{MRM}_{k,t,f}=\text{Sigmoid}(\mathbf{a}_k^T\mathbf{a}_{t,f}), &&
\end{flalign}
where $\mathbf{a}_k\in \mathbb{R}^{D\times 1}$ is the $D$-dimensional attractor of speaker $k$. The DAN is trained by minimizing the reconstruction loss for both separation and dereverberation,
\begin{flalign}
\label{eq:reconloss}
L_\text{r}=\sum_{k,t,f}(y_{t,f}\hat{m}^\text{MRM}_{k,t,f}-d_{k,t,f})^2. &&
\end{flalign}
The optimization leads to an embedding pattern that the vectors from the same speakers become more similar and those from different speakers become more discriminative. However, due to $y_{t,f}\neq \sum_{k}d_{k,t,f}$, Eq.(\ref{eq:reconloss}) may lead to performance degradation in clustering, which can be relieved by adding extra clustering losses (Section \ref{sec:cst_loss}).

\begin{figure}
	\centering
	\includegraphics[width=0.7\linewidth]{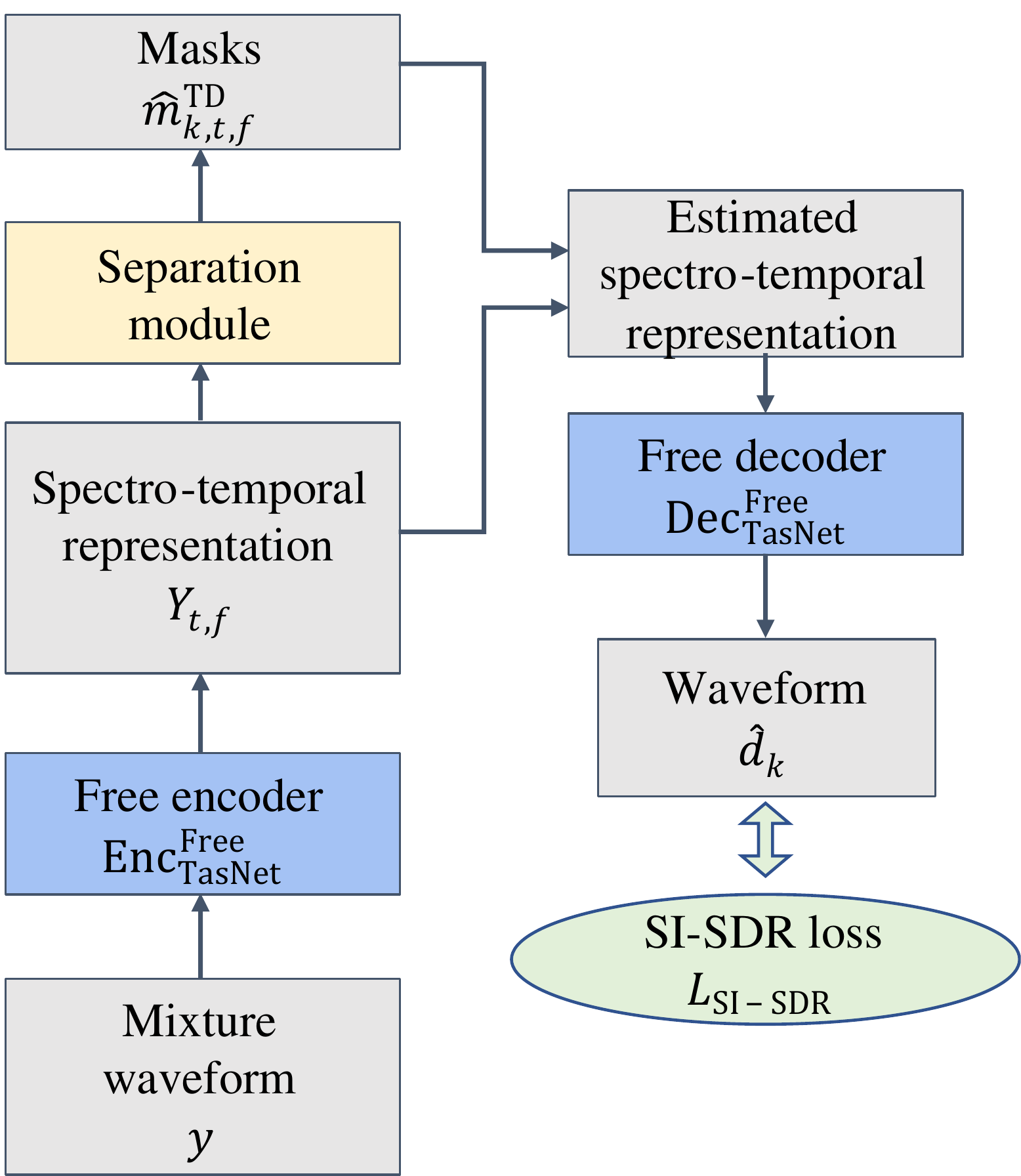}
	\caption{The architecture of Conv-TasNet, where the separation module generates masks $\{\hat{m}_{k,t,f}\}_{k,t,f}$ for a predefined number of speakers.}
	\label{fig:tasnet}
\end{figure}

\subsubsection{Conv-TasNet}

Conv-TasNet is a fully convolutional time-domain audio separation network, composed of a $1$-D convolutional encoder, a separation module and a $1$-D convolutional decoder. Multiple sequential TCN blocks with various dilation factors are stacked as the separation module. The fully convolutional architectures result in a small-sized model. As plotted in Fig.\ref{fig:tasnet}, the encoder encodes the input mixture signal,
\begin{flalign}
\{y_{t,f}\}_{t,f}=\text{Enc}^\text{Free}_\text{TasNet}(y), &&
\end{flalign}
where $\text{Enc}^\text{Free}_{\{\cdot\}}$ is a $1$-D time convolutional kernel and $y_{t,f}$ is the spectro-temporal representation. We use ``Free'' to indicate that the kernel parameters are learnable. The TCN-based separation module is trained to predict masks,
\begin{flalign}
\{\hat{m}^\text{TD}_{k,t,f}\}_{k,t,f}=\text{TCN}(\{y_{t,f}\}_{t,f}), &&
\end{flalign}
where $\hat{m}_{k,t,f}^\text{TD}$ is the estimated mask defined on the spectro-temporal representation. The decoder decodes the masked spectro-temporal representation and generates the enhanced waveforms,
\begin{flalign}
\{\hat{d}_k\}_k=\text{Dec}^\text{Free}_\text{TasNet}(\{y_{t,f}\hat{m}^\text{TD}_{k,t,f}\}_{k,t,f}), &&
\end{flalign}
where $\text{Dec}^\text{Free}_\text{TasNet}$ is a $1$-D time-domain kernel. Conv-TasNet uses utterance-level PIT (uPIT) to optimize the SI-SDR (\cite{Kolbaek2017MultitalkerSS}). 

\subsection{Time-domain deep attractor network}
\label{sec:TDhybrid}

The TD-DAN has a two-stream architecture composed of an SES for embedding modelling and an SDS for dereverberation and speaker extraction. We creatively separate the task into $2$ parts and jointly train the $2$ streams with a multi-task loss. We first describe the two-stream architecture together with the hybrid waveform encoders and then step into the fully time-domain encoders. 

\subsubsection{TD-DAN with hybrid encoders}

\begin{figure*}
	\centering
	\includegraphics[width=0.8\linewidth]{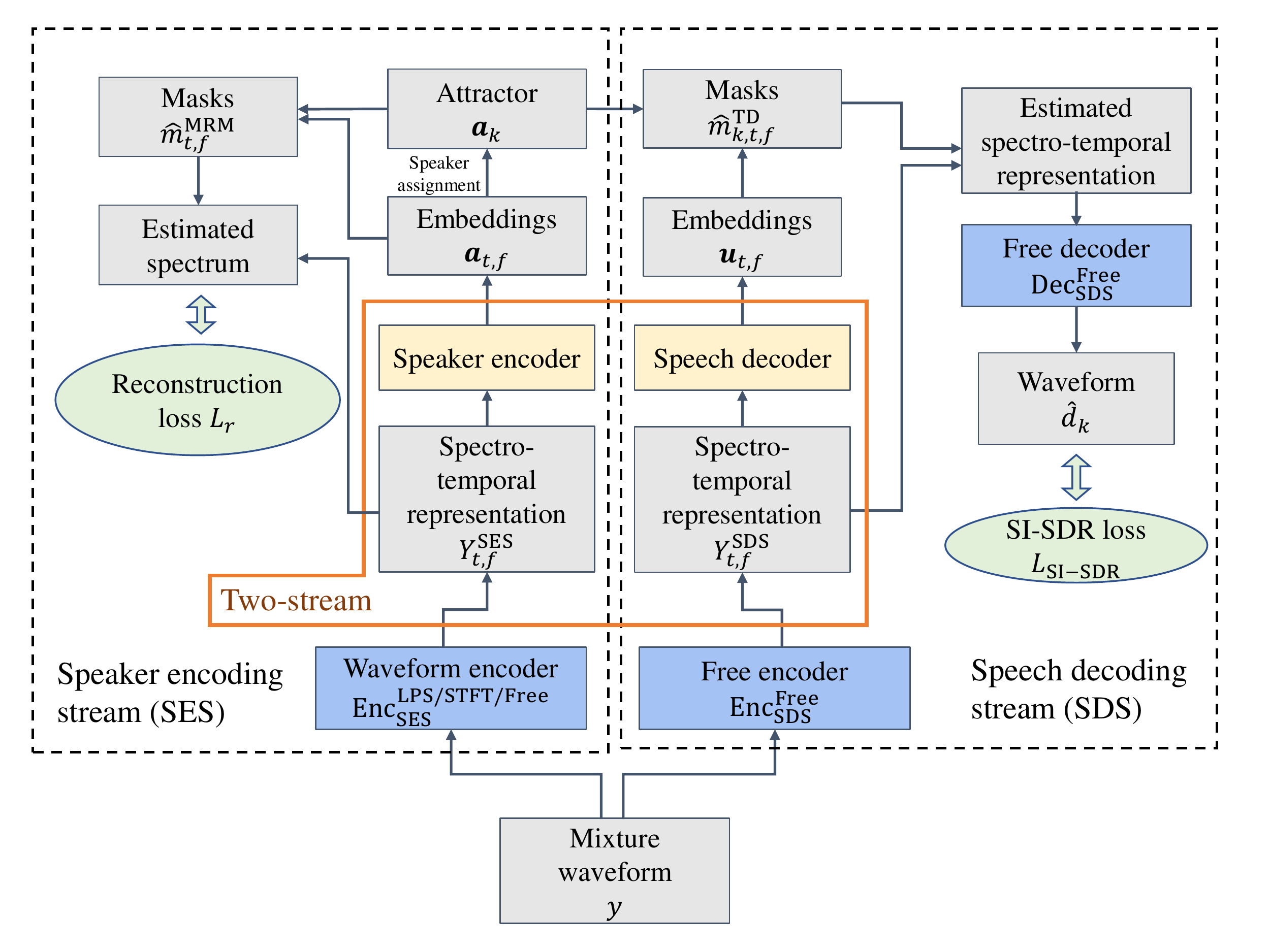}
	\caption{The architecture of TD-DAN, which is composed of an SES and an SDS. The waveform encoder of the SES can adopt frequency-domain LPS transform, time-domain stacked STFT kernels or free kernels.}
	\label{fig:tddan}
\end{figure*}

As plotted in Fig.\ref{fig:tddan}, the SES is similar to the DAN network, which accepts the LPS with $\text{Enc}_\text{SES}^\text{LPS}$ and calculates the masks with speaker embeddings and attractors. The whole feed-forward procedure follows Eqs.(\ref{eq:DAN})-(\ref{eq:reconloss}).

The SDS models the input signal with a $1$-D convolutional encoder and stacked TCNs,
\begin{flalign}
\label{eq:tddan_enc}
\{\mathbf{e}_{t,f}\}_{t,f}=\text{TCN}(\text{Enc}_\text{SDS}^\text{Free}(y)). &&
\end{flalign}
where $\mathbf{e}_{t,f}\in \mathbb{R}^{E\times 1}$ is the $E$-dimensional high-level representation. The SDS accepts the transformed attractor to calculate the masks and finally generates the dereverberated and separated signal,
\begin{flalign}
\label{eq:tddan_connect}
\hat{m}_{k,t,f}^\text{TD}=\text{ReLU}((\mathbf{a}_k)^T\mathbf{e}_{t,f}), &&
\end{flalign}
\begin{flalign}
\label{eq:tddan_dec}
\hat{d}_k=\text{Dec}_\text{SDS}^\text{Free}(\{\hat{m}_{k,t,f}^{TD}\mathbf{e}_{t,f}\}_{t,f}), &&
\end{flalign}
The model is trained to optimize a multi-task loss, 
\begin{flalign}
\label{eq:tddan_loss1}
L_\text{TD-DAN}=L_\text{SI-SDR}+\alpha_rL_\text{r}, &&
\end{flalign}
where $L_\text{SI-SDR}$ is calculated by comparing $d_k$ with $\hat{d}_k$, $\alpha_r$ is the loss balance factor.

This TD-DAN is with hybrid encoders because the SES is encoded by the STFT, while the SDS is encoded by a $1$-D convolutional encoder with free kernels. Nevertheless, it is regarded as a time-domain DAN since it is trained to predict waveforms directly.

\subsubsection{TD-DAN with fully time-domain encoders}
\label{sec:TDtime}

Here, we replace the waveform encoder $\text{Enc}_\text{SES}^\text{LPS}$ in the TD-DAN SES with time-domain convolutional kernels. The problem is the definition of the IBMs in the spectro-temporal representations, which are originally computed based on the spectrogram (Eq.(\ref{eq:attract_ibm})). The time-domain SES encoder $\text{Enc}_\text{SES}^\text{TD}$ encodes the mixture signal into $y_{t,f}$, formulated as
\begin{flalign}
\{y_{t,f}\}_{t,f}=\text{Enc}_\text{SES}^\text{TD}(y). &&
\end{flalign}
By setting the magnitude of the signal as $|y_{t,f}|$, its IBM is formulated similarly,
\begin{flalign}
\label{eq:TD_IBM}
m_{k,t,f}^\text{IBM}= \begin{cases}
0, \text{if $|\text{Enc}_\text{SES}^\text{TD}(d_{k,t,f})|\leqslant\sum_{q\neq k}|\text{Enc}_\text{SES}^\text{TD}(d_{q,t,f})|$} \\
1, \text{if $|\text{Enc}_\text{SES}^\text{TD}(d_{k,t,f})|>\sum_{q\neq k}|\text{Enc}_\text{SES}^\text{TD}(d_{q,t,f})|$}
\end{cases} &&
\end{flalign}

We introduce $2$ time-domain kernels, namely, the stacked time-domain STFT kernel and the free kernel :
\begin{itemize}
\item [1)] The stacked STFT encoder $\text{Enc}_\text{SES}^\text{STFT}$: The STFT is split into real and the imaginary parts with a stacked convolutional kernel expressed as follows,
\begin{flalign}
& K^{cos}_{f}[n]=w[n]cos(2\pi nf/N), && \\
& K^{sin}_{f}[n]=w[n]sin(2\pi nf/N), && \\
& \mathbf{K^{STFT}}=[\mathbf{K}^{cos}_{0},...,\mathbf{K}^{cos}_{F-1},\mathbf{K}^{cos}_{F},\mathbf{K}^{sin}_{1},...,\mathbf{K}^{sin}_{F-1}], &&
\end{flalign}
where $F$ usually equals $N/2$, columns of $\mathbf{K^{STFT}}$ are $1$-D convolutional kernels, $n$ is the sample index in a convolutional kernel of size $N$, $f=0,1,...,F$ is the kernel index corresponding to the frequency of the STFT, and $w$ is the pre-designed analysis window. This kernel is different from STFT since it stacks real and imaginary part of the spectrum, which can be conducted with real-valued convolutional operations.
\item [2)] The free convolutional encoder $\text{Enc}_\text{SES}^\text{Free}$: $1$-D convolutional kernel $\mathbf{K^{Free}}$ is trained together with the whole network.
\end{itemize}

The whole procedure with fully time-domain encoders follows Fig.\ref{fig:tddan}, where the attractor is obtained by masks defined by $\text{Enc}_\text{SES}^\text{TD}$ and is calculated by Eqs.(\ref{eq:attractor})-(\ref{eq:reconloss}); the dereverberation and separation are conducted following Eqs.(\ref{eq:tddan_enc})-(\ref{eq:tddan_dec}). Speech presence $v_{t,f}$ is obtained by a threshold of the magnitude of the spectro-temporal representations. The network is trained to optimize the multi-task loss (Eq.(\ref{eq:tddan_loss1})).

\subsection{Auxiliary clustering loss}
\label{sec:cst_loss}

\begin{figure}
	\centering
	\includegraphics[width=\linewidth]{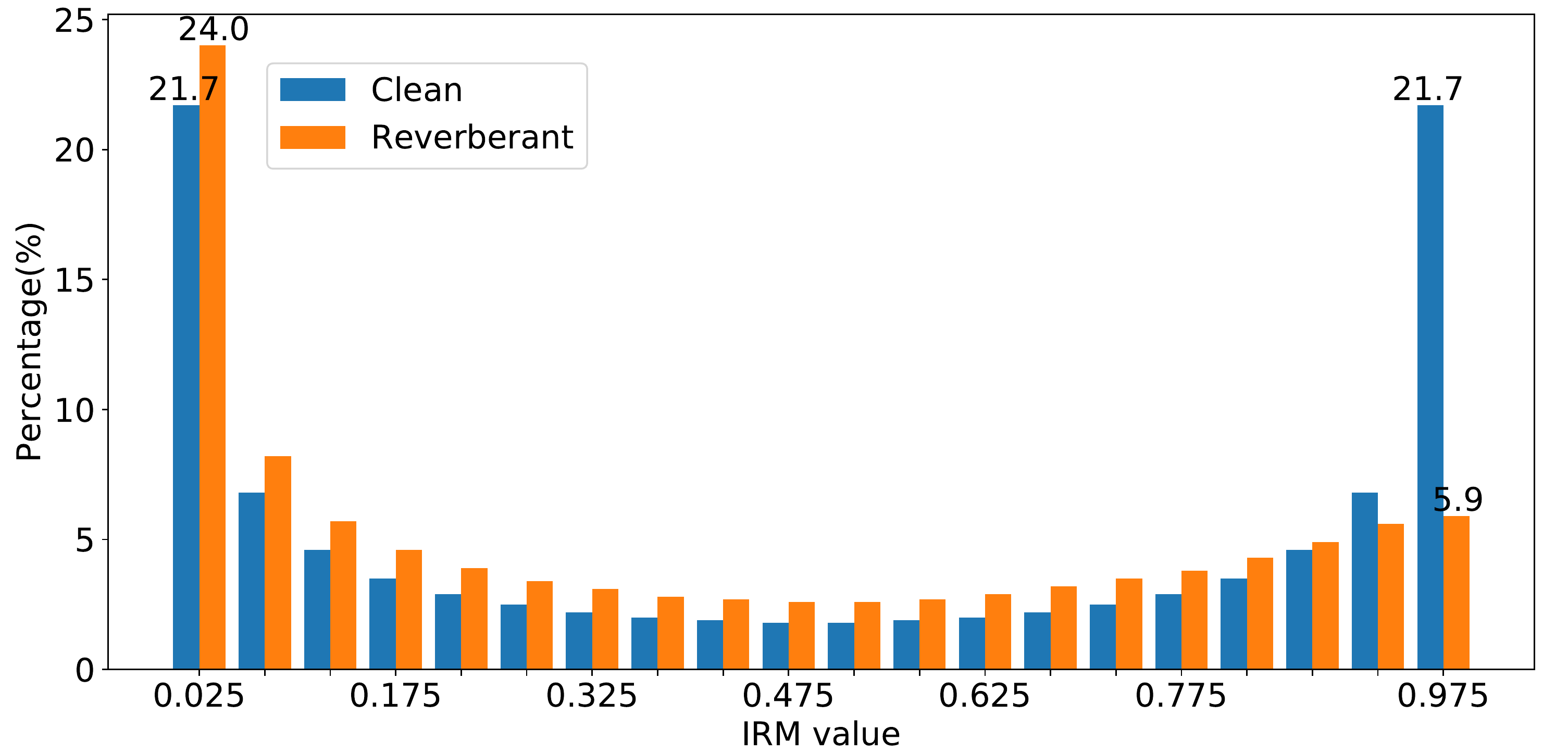}
	\caption{The histogram of the IRM calculated on the clean and reverberant multi-speaker mixtures. The clean mixture is mixed with early reflections, while the reverberant mixture is mixed with early and late reflections. The IRM is calculated with Eq.(\ref{eq:irm2}).}
	\label{fig:hist}
\end{figure}

The reconstruction loss (Eq.(\ref{eq:reconloss})) indicates that the mask will be near $1$ if the T-F bin embeddings are close to the speaker attractor, otherwise close to $0$. The sparsity assumption declares that the observed signal contains at most one source on each T-F bin, which ensures the clustering performance in the DAN since most embeddings are optimized so that they are close to some attractor to achieve binary-like masks. However, the reverberant signal may not follow the sparsity assumption. The distribution of the IRM in the mixture signal is plotted in Fig.\ref{fig:hist}. Notably, approximately $20\%$ T-F bins have an IRM value larger than $0.95$ in the mixture of early reflections, while in the reverberant signal, the percentage declines significantly to approximate $6\%$. The reason is that the IRM of early reflections is the ratio of the target early part to the interference early parts, while Eq.(\ref{eq:irm2}) is the ratio of the target early reflection to the target late reverberation, the interference early and late reverberation. The lack of high-value T-F masks causes difficulty in embedding clustering.

\begin{figure}
	\centering
	\includegraphics[width=\linewidth]{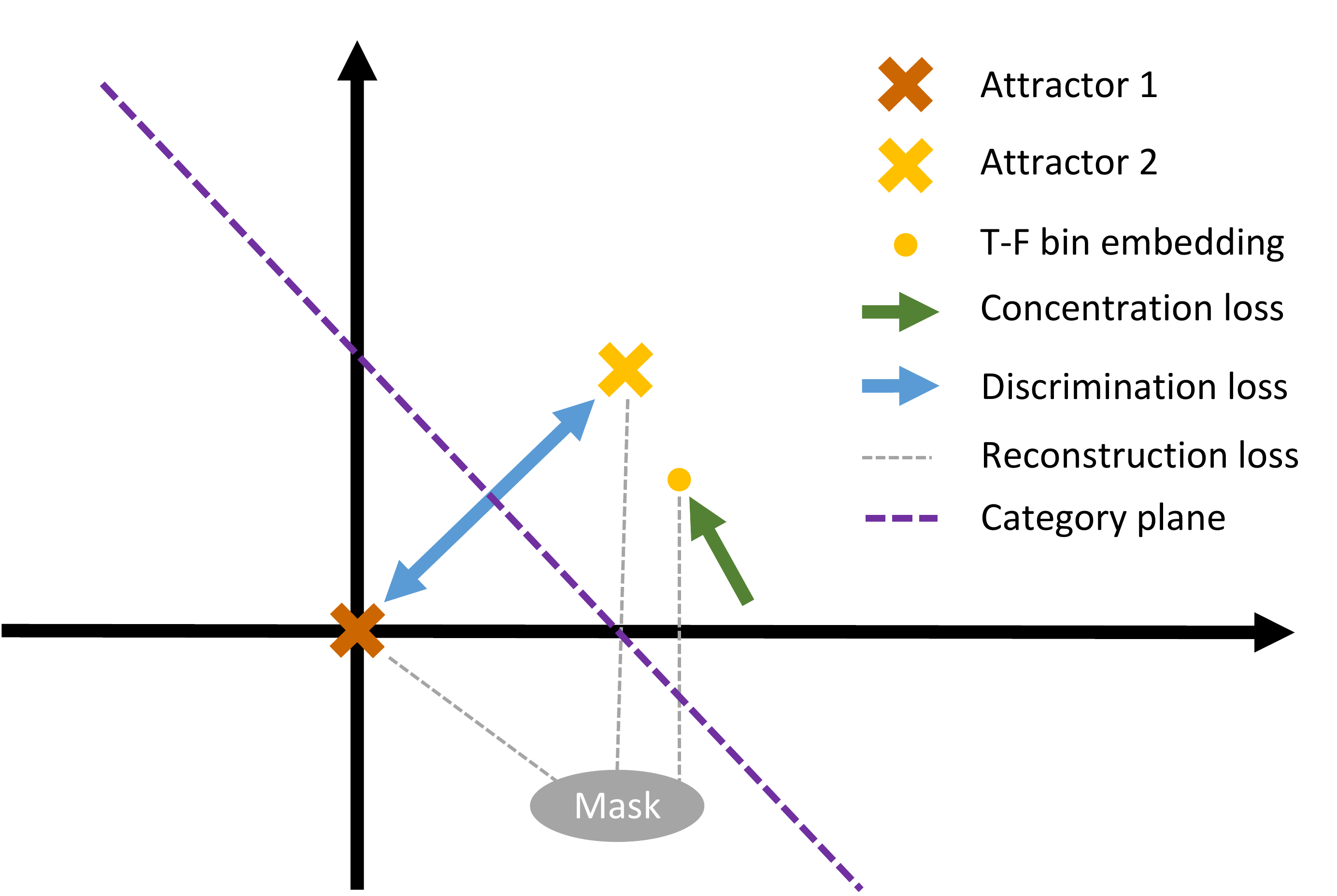}
	\caption{The diagram of the reconstruction loss and the clustering loss. The arrows of the clustering losses represent the optimization direction. The reconstruction loss with the mask constrains the attractors and the T-F embeddings. The category plane determines the dominated speaker on the embedding.}
	\label{fig:clusteringloss}
\end{figure}

To achieve a better clustering performance, we introduce the clustering loss, including the concentration loss and the discrimination loss. The concentration loss is designed for all DAN-based models,
\begin{flalign}
L_c=\sum_{k,t,f}||\mathbf{a}_k-m^\text{IBM}_{k,t,f}v_{t,f}\mathbf{a}_{t,f}||_2^2. &&
\end{flalign}
Its gradient is
\begin{flalign}
\frac{\partial L_c}{\mathbf{a}_{t,f}}=\begin{cases}
-2(1-\frac{1}{\sum_{t,f}m^\text{IBM}_{k,t,f}v_{t,f}})(\mathbf{a}_k-\mathbf{a}_{t,f}), \text{if $m^\text{IBM}_{k,t,f}v_{t,f}=1$} \\
0, \text{if $m^\text{IBM}_{k,t,f}v_{t,f}=0$}
\end{cases} &&
\end{flalign}
which enforces embedding $\mathbf{a}_{t,f}$ to be close to the attractor $\mathbf{a}_k$ when dominated by speaker $k$. 

Another inter-class discrimination loss maximizes the distance among different attractors,
\begin{flalign}
L_d=\text{max}(0,l_d^2-\sum_{k,q}^{k\neq q}||\mathbf{a}_{k}-\mathbf{a}_{q}||_2^2), &&
\end{flalign}
where the maximum distance $l_d$ is to avoid the network achieving a large distance by generating attractors with large norms. In fact, Eq.(\ref{eq:reconloss}) includes the optimization of discrimination, whereby the attractor distance will be enlarged if $m^\text{MRM}_{t,f}$ is close to $0$. The discrimination loss here is designed for free convolutional kernel $\mathbf{K^{Free}}$ in the SES where small $|\text{Enc}_\text{SES}^\text{Free}(y)|$ may result in small $L_r$ as well as the small inter-class distance. The training loss is updated to,
\begin{flalign}
\label{eq:tddan_loss2}
L_\text{TD-DAN}=L_\text{SI-SDR}+\alpha_rL_\text{r}+\alpha_cL_\text{c}+\alpha_dL_\text{d}, &&
\end{flalign}
where $\alpha_c$ and $\alpha_d$ are factors for the concentration and the discrimination losses.

Fig.\ref{fig:clusteringloss} presents a diagram to illustrate different losses. The intra-class concentration loss may conflict with Eq.(\ref{eq:reconloss}) to some degree, i.e., the loss pushes the embeddings concentrated around the attractors, which results in large-valued estimated masks and may lead to a suboptimal reconstruction loss, which was observed on DAN as described in Section \ref{sec:exp_dan}. For TD-DAN, the concentration loss might make the output of $\mathbf{K^{Free}}$ lose discrimination. But this problem will not occur on the TD-DAN with fixed SES encoders. Due to the proposed $2$-stream architecture, the clustering loss is applied on the SES branch, while the time-domain signal reconstruction is conducted on the SDS branch by using the attractors from the SES branch. The precision of the attractor plays an important role on the quality of the estimated signals. In practice, the joint optimization of the reconstruction and the concentration loss leads to a narrowed performance gap between the oracle and estimated attractors. The detailed experiments will be presented in Section \ref{sec:exp_cstloss}.

\section{Experimental configuration}

\subsection{Dataset}
The experiments were conducted on the Spatialized Multi-Speaker Wall Street Journal (SMS-WSJ) (\cite{Drude2019SMSWSJDP}). The performance was evaluated on the test sets of the datasets. We used K-means to obtain the attractor and $3$ measurement methods, SI-SDR, SDR and WER, to evaluate the performance. Without special annotation, the SI-SDR uses the corresponding learning target as the reference signal (early reflections), the SDR uses the clean signal as the reference signal, the signal was estimated by attractors from K-means clustering algorithm. The measurement methods are also discussed in detail in Section \ref{sec:learningtargets}. 

The SMS-WSJ dataset artificially spatialized and mixed utterances taken from WSJ. The dataset was split into the training, validation and test sets, which contained $33561$, $982$ and $1332$ utterances, respectively. The room impulse response (RIR) was randomly sampled with different room sizes, array centers, array rotation, and source positions. The sound decay time (T60) was sampled uniformly from $200$ ms to $500$ ms. The simulated $6$-channel audios contained early reflections ($<50 \text{ms}$), late reverberation ($>50 \text{ms}$), and white noise. The start sample of the room impulse response (RIR) was determined by finding the first sample which was larger than the maximum divided by ten. The end of the early part of the RIR was set to be $50$ ms after the start sample. The signal-to-interference ratio (SIR) and signal-to-noise ratio (SNR) for mixtures were randomly drawn from $-5$ dB to $5$ dB and from $20$ dB to $30$ dB, respectively. Moreover, we simulated a $3$-speaker dataset as a more challenging task, which used the same RIRs and utterance split as the SMS-WSJ dataset. The official automatic speech recognition (ASR) system was used to evaluate the word error rate (WER). In our experiments, we used only the first channel of the multi-channel signal. As demonstrated in Section \ref{sec:learningtargets}, the networks were trained to map the reverberant multi-speaker signal to early reflections.


\subsection{Training settings}

The experiments were conducted with Asteroid (\cite{Pariente2020Asteroid}), an audio source separation toolkit based on PyTorch (\cite{Paszke2017AutomaticDI}). We changed the DAN architecture from bi-directional long short-term memory (BLSTM) to TCN blocks, which allowed for fair comparison among different frameworks.

The two-stream TD-DAN was composed of the SES and the SDS, which adopted the architecture corresponding to the baseline DAN and Conv-TasNet, respectively. By following the hyper-parameter notations in (\cite{Luo2019ConvTasNetSI}), we list the architectures in Table \ref{tab:arch}, where all models repeated TCN blocks $4$ times. The power threshold was set to keep the top $15\%$ bins of the mixture spectrogram.

\begin{table}[htp]
	\caption{The model architectures with TCN hyper-parameter $B/H/P/X/R$, the embedding dimension of $\mathbf{a}_{t,f}/\mathbf{e}_{t,f}$ in the SES/SDS, and default loss factor $\alpha_{r/c/d}$.}
	\label{tab:arch}
	\begin{tabular}{cccc}
		\toprule
		Hyper-params. & DAN & Conv-TasNet & TD-DAN \\
		\midrule
		$B$    & $128$ & $128$  & $128$ \\
		$H$    & $512$ & $512$ & $512$    \\
		$P$    & $3$  & $3$ & $3$     \\
		$X$    & $4$  & $8$  & $8$  \\
		$R$    & $4$  & $4$ & $1\text{(SES)}+3\text{(SDS)}$    \\
		$\mathbf{a}$     & $20$ & $-$ & $20$   \\
		$\mathbf{e}$     & $20$ & $-$ & $20$     \\
		$\alpha_r$    & $1.0$ & $-$ & $1.0$     \\
		$\alpha_c$    & $0.05$ & $-$ & $1.0$      \\
		$\alpha_d$    & $0.0$ & $-$ & $0.0$      \\
		$l_d$    & $-$ & $-$ & $\sqrt{5}$      \\
		\bottomrule 
	\end{tabular}
\end{table}

We used the Adam optimizer (\cite{Kingma2015AdamAM}) with a learning rate starting from $10^{-3}$ and then halved if the best validation model was not found within $3$ epochs. The maximum number of epochs was set to $50$. The TD-DANs were trained with $4$-second segments and a batch size of $16$.

\section{Results and discussion}
\label{sec:res}

In this section, we will explore and discuss the performance of TD-DANs. Our goal is to improve the model's separation and dereverberation ability in a reverberant environment. The experiments will be presented in the following $4$ parts. First, we think that a reasonable learning target can ease the learning difficulty. Thus, choosing early reflections as learning targets was demonstrated by comparing different signals on the SMS-WSJ dataset. Second, DANs showed different characteristics when deployed under the reverberant environment. Thus, the model settings were adjusted in terms of the losses and power thresholds. Third, the TD-DAN model was explored by extending the DAN from the T-F domain to the time domain where the SES encoder and the clustering loss were studied in detail. Fourth, the TD-DAN was tested under the condition of a variable number of speakers and was compared with PIT-based multi-speaker separation paradigms. 

\subsection{Learning target comparison}
\label{sec:learningtargets}

The SMS-WSJ dataset provided the original clean signal, early reflections and reverberant signals for each mixture utterance. We simulated the anechoic signal additionally. These signals were chosen as the learning targets to demonstrate their difference. It was believed that the learning target should be close to the original clean signal and easy to be learned for the deep learning-based model. Thus, the comparison was conducted in $2$ aspects, signal measurement against the clean signal and training difficulty under the baseline Conv-TasNet.

\begin{table*}[b]
	\caption{Comparison of different learning targets in terms of SI-SDR, SDR, PESQ and STOI measurements, where the clean signal was used as the reference. The WER was measured with the official ASR baseline.}
	\label{tab:exp_learningtargets}
	\begin{tabular}{lccccc}
	\toprule
	Learning target & SI-SDR (dB) & SDR (dB)   & PESQ  & STOI  & WER (\%) \\ \midrule
	Anechoic        & $-15.04$ & $49.15$ & $\mathbf{4.53}$ & $\mathbf{1.00}$ &  $\mathbf{6.38}$   \\
	Early           & $-18.31$ & $\mathbf{49.46}$ & $2.35$ & $0.86$ &   $7.04$  \\
	Reverberation   & $-18.74$ & $14.86$ & $2.0$ & $0.83$ &  $8.17$ \\ \bottomrule 
	\end{tabular}
\end{table*}

Table \ref{tab:exp_learningtargets} compares different learning targets with the original clean signal. Following messages were obtained:
\begin{itemize}
\item SI-SDRs were low for all learning targets due to the convolution of the clean signal and the RIRs,
\begin{flalign}
	s_\text{reverberant/early/anechoic}=s_\text{clean}*\text{rir}, &&
\end{flalign}
where the convolution operator $*$ shifted and rescaled the signal, while the SI-SDR is sensitive to the shift. 
\item Source-to-distortion ratio (SDR, \cite{Vincent2006PerformanceMI}) allows the target signal located in a subspace spanned by the delayed version of clean signals. The filter length here was set to $512$ ($64$ ms with sample rate $8000$ Hz). The anechoic signals and early reflections could perfectly match the projected clean signal in the subspace. The reverberant signal, however, had a lower SDR since its RIR filter was longer than $200$ ms.  
\item The perceptual evaluation of subjective quality (PESQ, \cite{Rix2001PerceptualEO}) and the short-time objective intelligibility (STOI, \cite{Taal2011AnAF}) were calculated based on the power spectrum. On the one hand, the RIR length of the anechoic signal was short. The convolution operator mainly changed the phase in each frame. The power spectra of the anechoic and clean signals were nearly the same, resulting in the highest scores. On the other hand, late reverberation caused ``spectral smearing'' (\cite{Maciejewski2019WHAMRNA}), resulting in the lowest scores. 
\item WER is another objective measurement. Our acoustic model was trained using reverberant single-speaker signals following the baseline of SMS-WSJ. We found that the anechoic signal achieved the best performance, $0.66\%$ and $1.79\%$ lower than early and reverberant signals, respectively.
\end{itemize}

In conclusion, the early reflections could achieve a similar SDR and a slightly higher WER than the anechoic signal. The SI-SDR, PESQ and STOI were sensitive to distortions caused by the time-invariant filters. However, the deep learning-based model needed a loss function to learn the signal mapping. The SI-SDR was easy to implement. It represented the similarity between the estimated signal and the training target. Thus, we chose $3$ measurement methods: SI-SDR, SDR and WER. In the rest of the paper, without special annotations, the SI-SDR uses the corresponding learning target as the reference signal, the SDR uses the clean signal as the reference signal. In most experiments, we tested only the SI-SDRs to evaluate the performance of the models quickly.

\begin{table}[]
	\caption{The performance of Conv-TasNet by using different learning targets. The ``SI-SDR'' was calculated by comparing the estimated signal with the learning targets. The ``SDR'' was calculated by comparing the estimated signal with the clean signal. }
	\label{tab:exp_baseline} 
	\begin{tabular}{lccc}
	\toprule
	Learning target & SI-SDR (dB) & SDR (dB) & WER (\%) \\ \midrule
	Anechoic        & $5.25$ & $7.36$    &   $45.09$  \\
	Early           & $8.04$ & $\mathbf{9.39}$       &  $\mathbf{36.01}$   \\
	Reverberant     &   $\mathbf{9.28}$     &     $8.39$        &   $37.38$  \\ \bottomrule 
	\end{tabular}
\end{table}

Table \ref{tab:exp_baseline} indicates that learning the mask from the reverberant signal to the early reflection was a preferred choice. The early reflection made the estimated signals have high signal quality (SDR: $9.39$ dB) and relatively low ASR error (WER: $36.01\%$). Since the acoustic model (AM) was trained on the reverberant signals, the high WER indicated that time-domain mapping introduced much distortion, which was unseen for the AM. Early reflections were chosen as the learning target for models to perform both separation and dereverberation tasks.

\subsection{Exploring DANs in a reverberant environment}
\label{sec:exp_dan}

\begin{table}[]
	\caption{Experiments on DAN models with various hop size (ms) and power percentages (top $N\%$ T-F bins). The hop size was set as the half of the window size. The performance of the oracle attractor (Oracle) is also listed. All models employed $R(4)\times X(4)\times H(512)$ TCN blocks.}
	\label{tab:exp_dan} 
	\begin{tabular}{lcccc}
	\toprule
	\multirow{2}{*}{Model} & Hop size & Top $N\%$  & \multicolumn{2}{c}{SI-SDR (dB)}                     \\
	&  (ms)     &   bins  & \multicolumn{1}{l}{K-means} & Oracle \\
	\midrule
	Conv-TasNet      & -               & -              & \multicolumn{2}{c}{$6.58$}    \\
	DAN (w/o $L_c$)  & $16$            & $50$           & $6.51$            &  $6.96$ \\
	DAN (w/  $L_c$)  & $16$            & $50$           & $6.77$   &  $6.77$ \\ \midrule
	DAN (w/  $L_c$)  & $32$            & $50$           & $6.92$   &  $6.97$ \\
	DAN (w/  $L_c$)  & $64$            & $50$           & $6.22$            &  $6.23$ \\
	DAN (w/  $L_c$)  & $8 $            & $50$           & $6.03$            &  $6.05$ \\ \midrule
	DAN (w/  $L_c$)  & $32$            & $15$           & $\mathbf{6.95}$   &  $\mathbf{6.98}$ \\
	DAN (w/  $L_c$)  & $32$            & $90$           & $6.76$            &  $6.97$ \\ 
	\bottomrule 
	\end{tabular}
\end{table}

The DAN was evaluated to perform both separation and dereverberation tasks (Table \ref{tab:exp_dan}). The performance of the DAN surpassed that of the Conv-TasNet in a small-size model setting ($X=4$ here instead of $X=8$). Adding the concentration loss narrowed the performance gap between K-means clustering and oracle attractor. However, as we have stated in Section \ref{sec:cst_loss}, the concentration loss resulted in a suboptimal model with a lower signal measurement under oracle attractors (w/o $L_c$: $6.96$ dB vs w/ $L_c$: $6.77$ dB). 

\begin{figure*}
	\centering
	\includegraphics[width=1.0\linewidth]{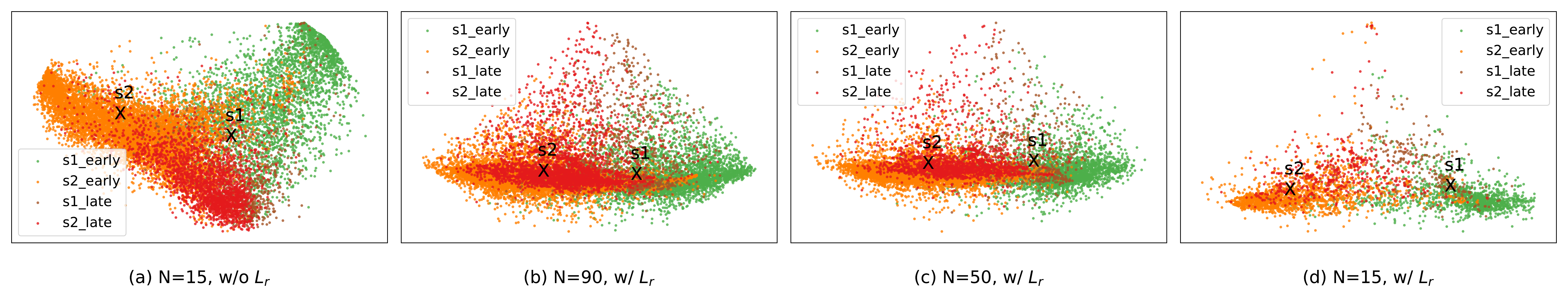}
	\caption{Visualization of embedding features from DAN models with different power thresholds (top N\% T-F bins) and $L_c=0.0/1.0$.}
	\label{fig:danembedding}
\end{figure*}

The window settings and power percentage were tuned to be a hop size of $32$ ms and the top $15\%$ T-F bins. The embedding features are visualized in Fig.\ref{fig:danembedding}. According to Figs.\ref{fig:danembedding}(a)-(b), the concentration loss concentrated the embeddings into a more compact pattern, resulting in higher SI-SDRs for the attractors calculated by K-means clustering. Since the late reverberation usually exhibited lower energy than the early reflections, lowering the power thresholds excluded the embeddings generated by late reverberation (Figs.\ref{fig:danembedding}(b)-(d)).  A more accurate attractor was obtained by aggregating more embeddings from the T-F bins dominated by early reflections.

\subsection{Exploring TD-DANs in a reverberant environment}
\label{sec:exp_tddan}

\subsubsection{Extending DAN to TD-DAN}

\begin{table}[]
	\caption{Experiments on TD-DAN models under different hop sizes, SES encoders and architectures. The hop size of the SES was set to the half of the window size. The default architecture settings for models were $R(4)\times X(8)\times H(512)$ TCN blocks. The DANs/TD-DANs were trained with a combination of the losses ($\alpha_r/\alpha_c/\alpha_d=1.0/1.0/0.0$ for the LPS and STFT encoders, $\alpha_r/\alpha_c/\alpha_d=1.0/1.0/1.0$ for the free encoders). The SES and SDS had $1$ and $3$ TCN blocks, respectively. TD-DAN ($1$-stream) merged the SES and SDS module into $1$ stream, whose input and output are the concatenated features of the ones of the SES and the SDS. }
	\label{tab:exp_tddan} 
	\resizebox{0.48\textwidth}{!}{
	\begin{tabular}{lccc}
	\toprule
	\multirow{2}{*}{Model}                                          & \multicolumn{2}{c}{SES} & \multirow{2}{*}{SI-SDR (dB)} \\
																	& Encoder & Hop size (ms) &                         \\
	\midrule
	Conv-TasNet ($X=4$)                                              & -       & -             & $6.58$          \\
	Conv-TasNet                                        & -       & -             & $\mathbf{8.04}$          \\
	DAN         ($X=4$)                                        & LPS     & $32$          & $6.95$                   \\
	DAN                                               & LPS     & $32$          & $6.83$                   \\
	\midrule
	TD-DAN                                                          & LPS     & $32$          & $7.97$                   \\
	TD-DAN                                                          & LPS     & $1 $          & $8.37$                   \\
	TD-DAN                                                          & LPS     & $2 $          & $8.44$                   \\
	TD-DAN                                                          & LPS     & $4 $          & $8.37$                   \\
	TD-DAN                                                          & STFT    & $2 $          & $\mathbf{8.69}$          \\
	TD-DAN                                                          & Free    & $2 $          & $8.08$                   \\
	\midrule
	\begin{tabular}[c]{@{}l@{}}TD-DAN\\ ($1$-stream)\end{tabular}     & STFT    & $1 $          & $8.22$                 \\
	\bottomrule                 
	\end{tabular}}
\end{table}

The $1$st part of Table \ref{tab:exp_tddan} displays the results of our baseline models, Conv-TasNet and DAN. The Conv-TasNet achieved an SI-SDR of $8.04$ dB, $1.21$ dB better than that of the DAN on the Eval. set. Compared with Table \ref{tab:exp_dan}, the performance of Conv-TasNet was vastly improved after increasing the convolutional layer number ($X=8$). The reason might be that the deep model benefit the time-domain modelling and had a larger reception field, which helped the model to conduct dereverberation and separation tasks. The DAN model did not exhibit performance improvement due to its large window size and the T-F domain representation.

The TD-DAN was designed following the architectures of DAN and TasNet. As listed in the $2$nd part of Table \ref{tab:exp_tddan}, the TD-DAN gave an SI-SDR of $7.97$ dB with the LPS encoder combined with the SDS, slightly lower than the SI-SDRs of Conv-TasNet. Since the deep embeddings from the SDS branch were from the time domain with a small hop size, the model could achieve better performance by eliminating the mismatch between the attractors and the time-domain embeddings. The TD-DAN could achieve an SI-SDR of $8.53$ dB by setting the hop size to $2$ ms and using STFT encoders. Predefined SES encoders were a preferred choice in the task as the free encoder achieved an SI-SDR of only $8.08$ dB.

As listed in the $3$rd part of Table \ref{tab:exp_tddan}, the TD-DAN was compared with the $1$-stream model. The $1$-stream TD-DAN accepted the concatenated features from the SES and SDS encoders and then processed the representation with the $1$-stream TCN model. The generated deep embeddings were split into SES and SDS parts. The SI-SDR of the $1$-stream model was $0.47$ dB lower than that of the best TD-DAN, implying the effectiveness of using $2$ separate modules for different embeddings.

\begin{figure*}
	\centering
	\includegraphics[width=1.0\linewidth]{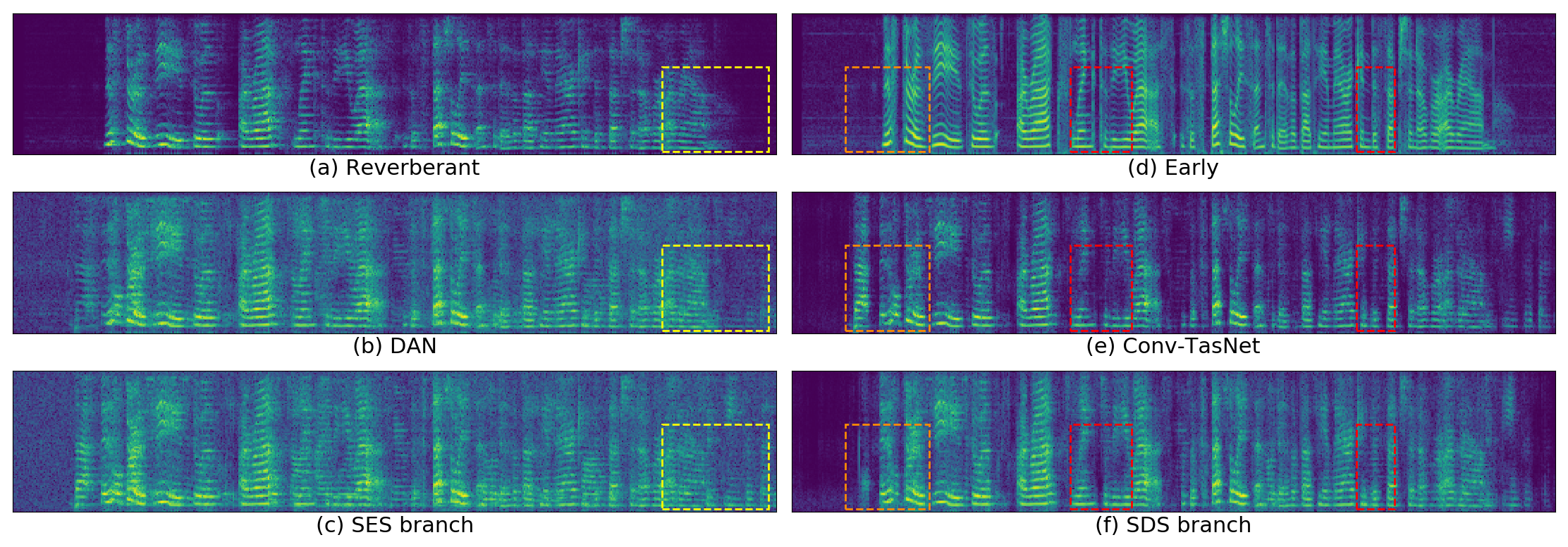}
	\caption{Enhanced spectra with different models under the 2-speaker condition. The enhanced spectra from (b) DAN model and (c) SES branch in TD-DAN exhibited low signal quality but they performed both the dereverberation and separation tasks (yellow boxes). Compared with (e) Conv-TasNet, (f) SDS branch in TD-DAN achieved a better performance in removing interference signals (orange boxes) and preserving the target signals (red boxes).}
	\label{fig:spectro}
\end{figure*}

Fig.\ref{fig:spectro} plots the enhanced STFT spectra estimated by different models. The DAN and the SES branch could perform dereverberation tasks (Figs.\ref{fig:spectro}(a)-(c)). However, the spectrum reconstruction exhibited lower signal quality than the early reflections and the one from the time-domain Conv-TasNet. The TD-DAN achieved better performance by removing more interference signals and preserving the target speech (Figs.\ref{fig:spectro}(d)-(f)).

\subsubsection{Analysis of the clustering loss}
\label{sec:exp_cstloss}


\begin{table}[]
	\caption{The SI-SDR (dB) results of TD-DAN models with STFT/free SES encoders.}
	\label{tab:exp_loss} 
	\begin{tabular}{ccccccc}
	\toprule
	\multirow{2}{*}{$L_r$} & \multirow{2}{*}{$L_c$} & \multirow{2}{*}{$L_d$} & \multicolumn{2}{c}{STFT} & \multicolumn{2}{c}{Free} \\
						&                     &                     & K-means       & Oracle      & K-means      & Oracle      \\
	\midrule
	\checkmark      &     \xmark &   \xmark   & $8.47$ & $\mathbf{8.95}$ & $6.86$ & $\mathbf{8.47}$ \\
	\checkmark      & \checkmark &   \xmark   & $\mathbf{8.69}$ & $8.94$ & $7.30$ & $8.23$ \\
	\xmark          & \checkmark & \checkmark & $7.74$ & $8.14$ & $7.93$ & $8.30$ \\
	\checkmark      & \checkmark & \checkmark & $7.86$ & $8.21$ & $\mathbf{8.08}$ & $8.42$ \\
	\bottomrule
	\end{tabular}
\end{table}

An ablation study was conducted to validate the effectiveness of the clustering loss with the stacked STFT and free SES encoders (Table \ref{tab:exp_loss}). For STFT encoders, the concentration loss helped the embedding much more concentrated, resulting in a smaller gap of the oracle attractors and the ones from K-means (Oracle: $0.48$ dB vs K-means: $0.25$ dB). The concentration loss made a little effect on the performance under the oracle attractors since the SDS branch only needed the estimated attractors. We observed a performance degradation by replacing the reconstruction loss with the discrimination loss ($L_c+L_r$: $8.69$ dB vs $L_c+L_d$: $7.74$ dB). The reconstruction loss here played the role of enlarging the inter-class distance implicitly by constraining the estimated masks. Yet the discrimination loss delivered a more straightforward way. It was thought that the reconstruction loss was preferred since it offered more detailed distance information with the reconstructed masks. The discrimination loss was unnecessary here as the reconstruction loss already provided considerable inter-class distance in our observation. Applying all the $3$ losses led to performance degradation since the discrimination loss might change the optimal pattern of the embeddings with the fixed STFT encoders. 

For free encoders, the concentration helped the clustering process. Meanwhile, it made the SES encoder generate representations less discriminative, which might explain performance degradation on the oracle attractors ($L_r$: $8.47$ dB vs $L_c+L_r$: $8.23$ dB). This phenomenon was not observed on the STFT encoder because the STFT weights were fixed and covered the whole frequencies. The discrimination loss delivered a narrow gap of the SI-SDRs between the oracle attractor and K-means. The reason was that the reconstruction loss could be lowered by generating small-valued spectro-temporal representations, as indicated in Eq.(\ref{eq:reconloss}). The discrimination loss and the concentration loss explicitly optimized the inter- and intra-class distance, forcing the network to generate embeddings easy to perform clustering. Combining all the $3$ losses offered a slight performance improvement, where both the clustering loss and the reconstruction loss assisted the free encoders in forming the pattern of the deep embeddings.

\subsection{Exploring TD-DANs with a variable number of speakers}

\begin{table*}[htp]
	\caption{Performance measurement of SI-SDR/SDR/WER under $1$-, $2$- and $3$-speaker conditions. ``1/2/3'' means that we trained $1$-, $2$- and $3$-speaker Conv-TasNet on the $1$-, $2$- and $3$-speaker dataset individually. }
	\label{tab:3spk_exp}
	\resizebox{0.98\textwidth}{!}{
	\begin{tabular}{lcccccccccc}
	\toprule
	\multirow{2}{*}{Model} & Training set & \multicolumn{3}{c}{1 speaker} & \multicolumn{3}{c}{2 speakers} & \multicolumn{3}{c}{3 speakers} \\
		& Speaker \#  &    SI-SDR(dB) & SDR(dB) & WER(\%)  & SI-SDR(dB) & SDR(dB) & WER(\%) & SI-SDR(dB) & SDR(dB) & WER(\%) \\
	\midrule
	DAN              & 2    & $12.75$ & $15.43$ & $11.06$ & $6.95$ & $8.42$ & $47.08$ & $-0.50$ & $0.05$ & $80.87$ \\
	DAN              & 1+2+3  & $12.93$ & $15.68$ & $10.62$ & $7.03$ & $8.24$ & $48.18$ & $3.02$  & $3.86$ & $76.94$ \\
	TD-DAN           & 2    & $14.25$ & $16.74$ & $11.38$ & $8.69$ & $9.90$ & $35.57$ & $0.20$  & $0.83$ & $77.56$ \\
	TD-DAN           & 1+2+3  & $\mathbf{14.64}$ & $\mathbf{17.08}$ & $\mathbf{9.00}$  & $\mathbf{8.95}$ & $\mathbf{10.22}$& $\mathbf{33.03}$ & $\mathbf{3.70}$  & $\mathbf{4.82}$ & $\mathbf{66.04}$ \\ 
	\midrule
	Conv-TasNet      & 1/2/3  & $\mathbf{14.53}$ & $\mathbf{16.89}$ & $9.38$  & $\mathbf{8.04}$ & $9.31$ & $\mathbf{36.01}$ & $\mathbf{3.23}$ & $\mathbf{4.29}$ & $70.96$ \\
	A2PIT            & 1+2+3  & $13.89$ & $16.52$ & $\mathbf{9.14}$  & $8.01$ & $\mathbf{9.33}$ & $36.30$ & $2.36$ & $3.34$ & $76.05$  \\
	ORPIT            & 1+2+3  & $13.64$ & $16.27$ & $9.20$ & $9.20$ & $9.10$ & $37.95$ &  $2.91$ & $4.26$ & $\mathbf{67.73}$ \\ \midrule
	Mixture          & -    & $11.75$ & $14.52$ & $8.90$  & $-0.84$& $-0.41$& $78.36$ & $-3.77$& $-3.37$& $91.46$ \\
	IRM(Eq.(\ref{eq:irm1})) & -  & - & - & - & $8.61$ & $10.25$ & $8.63$ & $7.20$ & $8.62$ & $9.21$ \\
	IRM(Eq.(\ref{eq:irm2})) & -  & $\mathbf{15.30}$ & $17.72$ & $7.77$ & $\mathbf{10.53}$ & $\mathbf{11.69}$ & $\mathbf{7.52}$ & $\mathbf{8.69}$ & $\mathbf{9.69}$ & $7.90$ \\
	WFM(Eq.(\ref{eq:wfm1})) & -  & - & - & - & $8.37$ & $9.84$ & $8.70$ & $6.94$ & $8.17$ & $9.16$ \\
	WFM(Eq.(\ref{eq:wfm2})) & -  & $15.19$ & $\mathbf{18.02}$ & $\mathbf{7.74}$ & $10.27$ & $11.37$ & $7.74$ & $8.44$ & $9.32$ & $\mathbf{7.79}$ \\
	\bottomrule
	\end{tabular}}
\end{table*}

The merit of the TD-DAN is that it can deal with mixture signals with variable numbers of speakers. To validate this feature, we further trained the DAN/TD-DAN on $2$ and $3$-speaker datasets. Besides, approximate $10\%$ samples were chosen as the $1$-speaker condition, i.e., the input and learning target were $1$-speaker reverberant signals and early reflections, respectively. The experiment results are listed in the $1$st part of Table \ref{tab:3spk_exp}. The DAN/TD-DAN only trained on the $2$-speaker dataset could deal with $1$-speaker reverberant signals and $3$-speaker mixture signal with SI-SDR gains of $1.00/2.50$ dB and $3.27/3.97$ dB compared with the reverberant input signals, respectively. After trained on the concatenated dataset, the DAN/TD-DAN achieved higher SI-SDR gains of $1.18/2.89$ dB, $7.87/9.79$ dB and $6.79/7.47$ dB on the $1$-, $2$- and $3$-speaker datasets, respectively.

We compared the TD-DAN with PIT-based models under the condition of a variable number of speakers, including individual Conv-TasNet trained on the 1-/2-/3-speaker datasets, trained with auxiliary autoencoding permutation invariant training (A2PIT, \cite{Luo2020SeparatingVN}) and trained with one-and-rest permutation invariant training (ORPIT, \cite{Takahashi2019RecursiveSS}). The output of the ORPIT was the single-speaker early reflection and the mixture of the residual reverberant signals. It needed $K$ iteration to estimate early reflections from $K$ speakers, while other models could output the estimated early reflections in $1$ pass. In most cases, the individually trained Conv-TasNet obtained the best performance. The ORPIT presented a large gap between the SI-SDR and the SDR and a lower WER. The reason might be that the signal shift might occur when the early reflection was obtained based on the estimated reverberant mixture. The low WER indicated that iterative separation could preserve more speech details than the Conv-TasNet on the $3$-speaker condition.

It was observed that the TD-DAN could achieve the best performance, surpassing Conv-TasNet/A2PIT/ORPIT by an SDR of $0.47/1.34/0.79$ dB on the $3$-speaker dataset. All models exhibited speech distortion, resulting in high WERs tested on the AM trained only on the reverberant signals. The TD-DAN model achieved the lowest WERs by preserving more speech cues on the spectrum, presented in Fig.\ref{fig:spectro}. 

The performance of ideal masks is listed in the $3$rd part of Table \ref{tab:3spk_exp}. The SI-SDR gap of $4.99$ dB between IRM (Eq.(\ref{eq:irm2})) and the TD-DAN on the $3$-speaker dataset indicates that performing multi-speaker separation and dereverberation remains a challenging task.

\section{Conclusion}

In this paper, we explored a framework of TD-DANs for speech separation tasks in a reverberant environment. We used different waveform encoders, including the LPS encoder, the stacked STFT and free convolutional kernels. The experimental results implied that the TD-DAN with the stacked STFT encoder achieved the best performance, surpassing the baseline Conv-TasNet and DAN model in terms of SI-SDR, SDR and WER on the 1-, 2- and 3-speaker dataset. We anticipate further exploring the TD-DAN architecture with the multi-channel information for better dereverberation and separation in future work.

\section*{Acknowledgment}
This work is partially supported by the Strategic Priority Research Program of Chinese Academy of Sciences (No. XDC08010300), the National Natural Science Foundation of China (Nos. 11590772, 11590774, 11590770, 11774380). 

\bibliographystyle{model5-names}

\bibliography{cas-refs}

\end{document}